\documentstyle[12pt,manuscript,aps,epsf]{revtex}  
\begin{document}
\newcommand{\p}{\partial}
\newcommand{\ls}{\left(}
\newcommand{\rs}{\right)}
\newcommand{\beq}{\begin{equation}}
\newcommand{\eeq}{\end{equation}}
\newcommand{\beqa}{\begin{eqnarray}}
\newcommand{\eeqa}{\end{eqnarray}}
\newcommand{\bdm}{\begin{displaymath}}
\newcommand{\edm}{\end{displaymath}}
\newcommand{\fps}{f_{\pi}^2 }
\newcommand{\mks}{m_{{\mathrm K}}^2 }
\newcommand{\ms}{m_{{\mathrm K}}^{*} }
\newcommand{\mk}{m_{{\mathrm K}} }
\newcommand{\msq}{m_{{\mathrm K}}^{*2} }
\newcommand{\rhos}{\rho_{\mathrm s} }
\newcommand{\rhob}{\rho_{\mathrm B} }
\draft
\title{Covariant kaon dynamics and kaon flow in heavy ion collisions} 
\author{Yu-Ming Zheng } 
\address{China Institute of Atomic Energy, P. O. Box 275(18), 
  Beijing 102413, P. R. China } 
\author{C. Fuchs, Amand Faessler, K. Shekhter } 
\address{Institut f\"ur Theoretische Physik der Universit\"at T\"ubingen, 
D-72076 T\"ubingen, Germany } 
\author{Yu-Peng Yan, Chinorat Kobdaj } 
\address{School of Physics, Suranarre University of Technology,  
 Nakhon Ratchasima 30000, Thailand }
\maketitle
\begin{abstract}
The influence of the chiral mean field on the $K^+$ transverse 
flow in heavy ion collisions at SIS energy is investigated 
within covariant kaon dynamics. For the kaon mesons inside the 
nuclear medium a quasi-particle picture including 
scalar and vector fields is adopted and compared to the standard 
treatment with a static potential. It is confirmed that a 
Lorentz force from spatial component of the vector field provides 
an important contribution to the in-medium kaon dynamics and 
strongly counterbalances the influence of the vector potential 
on the $K^+$ in-plane flow. The FOPI data can 
be reasonably described using in-medium kaon potentials 
based on effective chiral models. 
The information on the in-medium $K^+$ potential extracted 
from kaon flow is consistent with the knowledge 
from other sources.  
\pacs{{\bf PACS} number: 25.75.+r \\
{\it Keywords:} Covariant kaon dynamics, kaon flow, 
kaon mean field, heavy ion collision }
\end{abstract}
\section{Introduction}
Properties of kaons in dense hadronic matter are important for a better 
understanding of both, a possible restoration of chiral symmetry 
in dense hadronic matter and the properties of nuclear matter 
at high densities. It is known from chiral models that the 
kaon mean field is related to chiral symmetry 
breaking \cite{kaplan86}. The in-medium effects give rise to an 
attractive scalar potential inside the nuclear medium which is 
in mean field approximation proportional to  
the kaon-nucleon Sigma term $\Sigma_{\mathrm{KN}}$. A second part of 
the mean field originates from the interaction with vector mesons 
\cite{kaplan86,brown96,schaffner97,Waas,lutz}. The vector potential is 
repulsive for kaons $K^+$ and, due to G-parity conservation, attractive 
for antikaons $K^-$. A strong attractive potential for antikaons may 
also favor $K^-$ condensation at high nuclear densities and thus modifies 
the properties of neutron stars \cite{li97b}. 

There have been extensive experimental efforts to search for the 
kaon in-medium properties in heavy ion collisions, in particular 
at intermediate energies \cite{fopi95,kaos,menzel00,sturm01,fopi97}. 
Corresponding 
transport calculations revealed significant evidence for in-medium 
modifications of the kaon properties during the course of such 
reactions \cite{ko95,ko97a,giessen,nantes97,nantes02,wang97,wang98,fuchs98,fuchs01,zheng02}. 
The picture was recently complemented by the measurements of 
$K^+$ production in proton-nucleus reactions at 
COSY-ANKE \cite{anke}. While $A+A$ reactions test 
the kaon dynamics also at supra-normal nuclear densities, 
$p+A$ reactions can not exceed saturation density. However, 
in the latter case the reaction dynamics is theoretically 
much easier to handle (due to much less secondary scatterings)  
and therefore the interpretation of the data is more 
straightforward. From the $p+A$  data the existence of a repulsive 
in-medium $K^+$ potential of about 20 MeV at saturation density 
($U_K (\rho_0) \simeq 20$ MeV) was derived \cite{anke}.

In $A+A$ reactions the transverse flow of $K^+$ mesons is one of specially  
attractive observables. It was proposed by Ko and Li \cite{ko95} 
that the kaon flow pattern at the final stage can be used as 
a sensitive probe for the kaon potential in a nuclear medium.  
Other theoretical studies predicted similar features for the 
kaon flow \cite{giessen,wang98}. As pointed out by 
Fuchs at al. \cite{fuchs98}, however, the kaon dynamics used in the 
above investigations is non-covariant, i.e. it is based on 
a static potential like force. The 
Lorentz-force like contribution from a typical relativistic 
scalar-vector type structure is missing. However, this contribution 
strongly counterbalances the influence of the vector potential 
on the $K^+$ in-plane flow which makes it more difficult 
to draw definite conclusions from transverse flow pattern.

In the meantime the FOPI collaboration \cite{fopi99} published new 
data of $K^+$ side flow with improved precision. These open 
the possibility to study this problem in more detail. 
In this work we apply 
the covariant kaon dynamics to describe the kaon in-medium 
properties, discuss its influence on the $K^+$ in-plane flow 
in heavy ion collisions, and compare to 
the standard treatment with a static potential. Our studies 
show that a relatively strong repulsive $K^+$ potential as 
proposed by Brown $\&$ Rho can 
reproduce the new FOPI data in the covariant kaon   
dynamics quite well and the strength of the potential is 
roughly consistent with recent information obtained from 
$p+A$ scattering \cite{anke}. 
\section{The model}
Due to its relativistic origin, the kaon mean field has a typical 
relativistic scalar--vector type structure. For the nucleons such 
a structure is well known from Quantum Hadron Dynamics \cite{serot88}.  
This decomposition of the mean field is most naturally 
expressed by an absorption of the scalar and vector parts into 
effective masses and momenta, respectively, leading to 
a formalism of quasi-free particles inside the nuclear medium
\cite{serot88}. 

From the chiral Lagrangian the field equations for 
the $K^\pm$--mesons are derived from the 
Euler-Lagrange equations \cite{ko95} 
\beq
\left[ \partial_\mu \partial^\mu \pm \frac{3i}{4f_{\pi}^{*2}} j_\mu \partial^\mu 
+ \left( \mks - \frac{\Sigma_{\mathrm{KN}}}{f_{\pi}^{*2}} \rhos \right) 
\right] \phi_{\mathrm{K^\pm}} (x) = 0
\quad .
\label{kg1}
\eeq
Here the mean field approximation has already been applied. 
In Eq. (\ref{kg1}) $ \rhos$ is the baryon scalar density, 
$j_\mu$ the baryon four-vector current, 
$f_{\pi}^{*}$ the in-medium pion decay constant. 
Introducing the kaonic vector potential 
\beq 
V_\mu = \frac{3}{8f_{\pi }^{*2}} j_\mu , 
\label{vpot1}
\eeq
Eq. (\ref{kg1}) can be rewritten in the form \cite{fuchs98}
\beq
\left[ \left( \partial_\mu \pm i V_\mu \right)^2  + \msq \right] 
\phi_{\mathrm{K^\pm}} (x) = 0~. 
\label{kg2a}
\eeq
Thus, the vector field is introduced by minimal coupling 
into the Klein-Gordon equation. The effective mass $\ms$ of 
the kaon is then given by \cite{schaffner97,fuchs98,fuchs01,zheng02,zheng02a} 
\beq
\ms = \sqrt{ \mks - \frac{\Sigma_{\mathrm{KN}}}{f_{\pi }^{*2}} \rhos 
     + V_\mu V^\mu }
\quad . 
\label{mstar1}
\eeq
Due to the bosonic character, the coupling of the scalar 
field to the mass term is no longer linear as for the 
baryons but quadratic and contains an additional contribution 
originating from the vector field. The effective quasi-particle 
mass defined by Eq. (\ref{mstar1}) is a Lorentz scalar and 
is equal for $K^+$ and $K^-$. 
In nuclear matter at rest the spatial components of the 
vector potential vanish, i.e. ${\bf V} = 0$, and 
Eqs. (\ref{kg2a}) reduce to the expression already 
given in Ref. \cite{ko95}. 
However, Eqs. (\ref{kg2a}) generally account for the 
correct Lorentz properties which are not obvious from the 
standard treatment of the kaon 
mean field \cite{ko95,ko97a,giessen,li97b}. 

The covariant equations of motion are obtained in the classical 
(testparticle) limit from the relativistic transport 
equation for the kaons which can be derived from Eqs. 
(\ref{kg2a}). They are analogous to the 
corresponding relativistic equations for baryons and 
read \cite{fuchs98}
\beqa
\frac{ d  q^\mu}{d\tau} = \frac{k^{*\mu}}{\ms}
\quad , \quad 
\frac{ d  k^{*\mu}}{d\tau} = \frac{k^{*}_{\nu}}{\ms} F^{\mu\nu} 
+\partial^\mu \ms
\quad . 
\label{como}
\eeqa
Here $q^\mu = (t,{\bf q})$ are the coordinates in Minkowski space 
and $F^{\mu\nu} = \partial^\mu  V^\nu - \partial^\nu  V^\mu $ is the 
field strength tensor for $K^+$. For $K^-$ where the vector field 
changes sign. The equation of motion are identical, however, 
$F^{\mu\nu}$ has to be replaced by $-F^{\mu\nu}$. 
The structure of Eqs. (\ref{como}) may become 
more transparent considering only the spatial components
\beq
\frac{d {\bf k^*}}{d t} = - \frac{\ms}{E^*} 
\frac{\partial \ms }{\partial {\bf q}} \mp 
\frac{\partial V^0 }{\partial {\bf q}} 
\pm \frac{{\bf k}^*}{E^*} \times 
\left( \frac{\partial}{\partial {\bf q}} \times {\bf V} \right)
\label{lorentz}
\eeq
where the upper (lower) signs refer to $K^+$ ( $K^-$). 
The term proportional to the spatial component of the vector 
potential gives rise to a momentum dependence 
which can be attributed to a Lorentz force, i.e. 
the last term in Eq. (\ref{lorentz}). Such a velocity dependent 
$({\bf v} = {\bf k}^* / E^* )$ Lorentz force 
is a genuine feature of relativistic dynamics as soon 
as a vector field is involved. 

If the equations of motion are, however, derived from a static 
potential 
\beqa
 U(\rho,{\bf k}) & = & \omega(\rho, {\bf k}) - \omega_0 ({\bf k}) \nonumber \\
& = & \sqrt{{\bf k}^2 +  \mks - \frac{\Sigma_{\mathrm{KN}}}{f_{\pi}^{*2}} \rhos 
     + V_{0}^2 } \pm V_0 - \sqrt{{\bf k}^2 +  \mks }
\label{pot}
\eeqa
as given in Refs. \cite{li97b,ko95,ko97a,giessen,wang98}, 
the Lorentz-force (LF) like contribution is missing. 
Non-covariant treatments are formulated 
in terms of canonical momenta $k$ instead of kinetic 
momenta $k^*$ and then the equations of motion read
\begin{equation}
\frac{d {\bf k}}{d t} = - \frac{\ms}{E^*} 
\frac{\partial \ms }{\partial {\bf q}} \mp 
\frac{\partial V^0 }{\partial {\bf q}} 
\pm {\bf v}_{i} \frac{\partial  {\bf V}_{i}}{\partial {\bf q}} 
\quad ,
\label{lorentz2}
\end{equation}
with ${\bf v} = {\bf k}^* / E^*$ the kaon velocity. 

Following Brown and Rho \cite{brown96}, we use 
$ \Sigma_{\mathrm{KN}}$ = 450 MeV, 
$f_{\pi}^{*2}$ = 0.6$f_{\pi}^{2}$ for the vector field 
and $f_{\pi}^{*2}$ = $f_{\pi}^{2}$ for the scalar part 
given by $-\Sigma_{\mathrm{KN}}/f_{\pi}^{*2}\rhos $. This 
accounts for the fact that the enhancement 
of the scalar part using $f_{\pi}^{*2}$ is compensated 
by higher order corrections in the chiral expansion 
\cite{brown96,Waas}. This parameterization is denoted 
as Brown $\&$ Rho parameterization (BRP), which has 
already been used in our previous investigations 
\cite{fuchs98,fuchs01,zheng02,zheng02a}. 
For comparison also a weaker potential with 
$\Sigma_{\mathrm{KN}}$=350 MeV and 
$f_{\pi}^{*2} = f_{\pi}^{2}$ is applied. This parameterization 
was  originally used in ref. \cite{ko95}  
and is called the Ko and Li parameterization 
(KLP) in the following. 
The baryon dynamics are treated within the framework 
of Quantum Molecular Dynamics (QMD). For the nuclear 
forces we use the standard momentum dependent 
Skyrme interactions corresponding to a soft/hard EOS 
(K=200/380 MeV). For the determination of the kaon 
mean field we adopt the corresponding covariant 
scalar--vector description of the non-linear 
$\sigma\omega$ model. Here we use the parameterizations 
of \cite{ko97a,fang94} which correspond to identical 
soft/hard nuclear EOSs. 

The potentials given in Eq. (\ref{pot})  
at zero momentum by the BRP and KLP 
are shown in Fig. 1, where one also shows 
the kaon potential determined from 
the kaon-nucleon scattering length using the impulse 
approximation (IA) \cite{brown94}. In IA the energy 
$\omega(\rho,{\bf k})$, Eq. (\ref{pot}), is 
given as follows: 
\beq
\omega(\rho,{\bf k}) = 
\sqrt{{\bf k}^2 + \mks - 4\pi (1+\frac{m_K}{m_N})\bar{a}_{\mathrm{KN}} \rho }~~ , 
\label{potIA}
\eeq
where $m_N$ is the nucleon mass and  
$\bar{a}_{\mathrm{KN}} \approx $ -0.255 fm 
is the isospin-averaged kaon-nucleon 
scattering length in free space \cite{barnes94}. 
It is seen from Fig. 1 that the potentials 
predicted by BRP with the soft equation of 
state (solid curve) and by KLP with soft EOS 
(dashed line) are slightly stronger than that ones 
with the hard EOS (see the dot curve and  
long-dashed line, respectively). Up to saturation 
density the impulse approximation and the BR 
potential almost coincide ($U_K(\rho_0)\simeq 30$ MeV) 
but at supra-normal densities the BR potential 
rises much steeper than the IA. The KL potential 
($U_K(\rho_0)\simeq 5$ MeV, dashed line), 
on the other hand, is much weaker than those given 
by both, BRP and IA. 

The  $K^+$ creation mechanism is treated as described 
in refs. \cite{fuchs01,zheng02,zheng02a} 
where one uses the improved cross section of 
refs. \cite{sibirtsev95,tsushima99} for the baryon 
induced $K^+$ creation channels and of 
refs. \cite{tuebingen1} for the pion induced channels. 
The kaon production is treated perturbatively 
and does generally not affect the reaction dynamics 
\cite{ko97a,fuchs01,zheng02,zheng02a,fang94}. 
The shift of the production thresholds 
of the kaons by the in-medium potentials are taken into 
account as described in \cite{fuchs01,zheng02,zheng02a}. 
In order to account for energy-momentum conservation 
it is useful to formulate the mass shell condition, 
Eq. (\ref{kg2a}), in terms of the canonical momenta 
\beqa
0= k_{\mu}^{*2} - \msq = k_{\mu}^{2} - \mks - 2\mk U_{\rm opt} 
\label{mass2}
\eeqa
with
\beq
U_{\rm opt}(\rho ,{\bf k}) 
=  -\Sigma_S + \frac{1}{\mk} k_{\mu} V^{\mu}  
+ \frac{\Sigma_S^2 - V_{\mu}^2}{2\mk}  ~.
\label{uopt}
\end{equation}
Here we introduced the total scalar kaon self-energy 
$\Sigma_S = \ms - m_{{\mathrm K}}$. Since $U_{\rm opt}$ is a Lorentz scalar 
it can also be absorbed into an effective mass 
\beq
{\tilde m}_{\mathrm K} (\rho ,{\bf k}) 
= \sqrt{ \mks + 2\mk U_{\rm opt}(\rho ,{\bf k}) }
\label{effmass}
\eeq
which sets the canonical momenta on the mass-shell
\beqa
0= k_{\mu}^{*2} -\msq =  k_{\mu}^{2} - {\tilde m}_{\mathrm K}^{2}
\quad .
\label{mass3}
\eeqa
By definition ${\tilde m}_{\mathrm K}$ is a scalar but in contrast 
to $\ms$ who's analog is the Dirac mass in the case of nucleons 
  ${\tilde m}_{\mathrm K}$ absorbs the full optical potential 
and corresponds at zero momentum to the energy $\omega$. 
The threshold condition for $K^+$ production in baryon induced 
reactions reads then 
\beq
\sqrt{s} \ge {\tilde m}_B + {\tilde m}_Y + {\tilde m}_K
\label{tresh}
\eeq
with $\sqrt{s}$ the center--of--mass energy of the colliding baryons. 
The momenta of the outgoing particles are distributed according 
to the 3--body phase space
\beq
d\Phi _{3}(\sqrt{s},{\tilde m}_{B},{\tilde m}_Y, {\tilde m}_{K}) = 
d\Phi_{2}(\sqrt{s},{\tilde m}_{B},M)dM^{2}
\Phi_{2}(M,{\tilde m}_Y, {\tilde m}_{K})
\quad .
\label{phase4}
\eeq
The two-body phase space in Eq.(\ref{phase4}) has the form 
\begin{eqnarray}
\Phi _{2}(\sqrt{s},m_{1},m_{2})
=\frac{\pi p^{*}(\sqrt{s},m_{1},m_{2})}{\sqrt{%
s}}  \label{II.11}
\end{eqnarray}
where 
\begin{eqnarray}
p^{*}(\sqrt{s},m_{1},m_{2})=\frac{\sqrt{%
(s-(m_{1}+m_{2})^{2})(s-(m_{1}-m_{2})^{2})}}{2\sqrt{s}}  
\label{phase5}
\end{eqnarray}
is the momentum of the particles $1$ and $2$ in the c. m. frame.
Eq. (\ref{phase4}) corresponds to a distribution of the particle 
momenta according to an isotropic 3--body phase space. However, 
in \cite{li97b} a parameterization of the form 
\beq
d\Phi _{3}(\sqrt{s},{\tilde m}_{B},{\tilde m}_Y, {\tilde m}_{K}) = 
dW_{K}(\sqrt{s},{\tilde m}_{B},{\tilde m}_Y,M_K)dM^{2}_K
\Phi_{2}(\sqrt{s}-M_K,{\tilde m}_Y, {\tilde m}_{B})
\label{phase6}
\eeq
has been suggested where the kaon momentum $p$ is distributed 
according to 
\beq
dW_K \simeq \left(\frac{p}{p_{\rm max}}\right)^3 
\left(1-\frac{p}{p_{\rm max}}\right)^2 ~~, 
\eeq 
with $p_{\rm max}=p^*(\sqrt{s},{\tilde m}_{B}+{\tilde m}_Y, {\tilde m}_K )$ 
the maximal kaon momentum in the $BB$ c.m. frame. 
$M_K = \sqrt{p^2 +{\tilde m}_{K}^2}$ in Eq. (\ref{phase6}). 
The parameterization of Eq. (\ref{phase6}) has been motivated 
by analyzing corresponding $pp\rightarrow p\Lambda K^+$ data \cite{li97b} 
and shifts the kaon spectrum to lower momenta compared 
to an ideal 3--body phase space. 
The optical potentials of the baryons which enter via 
${\tilde m}_B,~{\tilde m}_Y$ are taken from the soft/hard EOS 
versions of the $\sigma\omega$ model \cite{ko97a,fang94}. 
The hyperon fields are thereby scaled according to SU(3) symmetry 
$U_{\rm opt}^Y = \frac{2}{3} U_{\rm opt}^B$. Since the ${\tilde m}$'s 
depend on the final state momenta the determination of 
$d\Phi _{3}$ is a self-consistency problem which is solved 
by iteration. The same procedure is applied to the 2-body-phase-space 
in pion induced reactions. The rescattering 
of the $K^+$ mesons with baryons and the Coulomb  
interaction are taken into account. 
The electromagnetic interaction is treated analogously 
to the strong interaction, i.e. 
by adding $F^{\mu\nu}_{{\rm el }} 
= \partial^\mu  A^\nu - \partial^\nu  A^\mu $ given by the 
electromagnetic vector potential to the Eq. (\ref{como}). 

\section{Results}
In order to study first the influence of the 3--body phase space 
in $BB$ collisions we consider in Fig. 2 $C+C$ collisions at 2.0 AGeV. 
The KaoS Collaboration has measured inclusive $K^+$ spectra at various 
c.m. angles with high precision. Corresponding calculations are performed 
with the soft EOS including the $K^+$ in-medium potential BRP and final state 
interactions by rescattering processes. 
In this scenario the excitation function of the 
kaon multiplicities measured by KaoS \cite{sturm01} is well reproduced 
over a large energy range from 0.8 to 2 AGeV \cite{fuchs01}. Without 
in-medium potentials the total multiplicities are 
significantly overpredicted. Looking to the spectra in more detail 
we find that an isotropic 3-body phase space in the 
$BB\rightarrow BYK^+$ channel 
shifts the spectra to too high momenta. The empirical parameterization 
of the 3-body phase space suggested in ref. \cite{li97b}, Eq. (\ref{phase5}), 
improves the situation but is still not fully sufficient to account for the 
angular asymmetry which is seen in the KaoS data \cite{sturm02}. 
A relatively good agreement can be achieved introducing 
an empirical angular dependence 
$d\sigma\propto (1+a\cos^2\theta_{\rm c.m.})d\cos\theta_{\rm c.m.} $ in 
the elementary cross sections.  
Following the data analysis of \cite{sturm02} an asymmetry parameter $a=1.2$ 
leads to slightly forward/backward peaked elementary 
$BB\rightarrow BYK^+$ cross sections  and the corresponding 
spectra shown in Fig. \ref{spec_fig} are then well reproduced by 
the transport calculations. 

In order to investigate the influence of covariant dynamics on the  
$K^+$ in-plane flow we consider the 1.93 AGeV 
$^{58}$Ni + $^{58}$Ni collisions at impact 
parameter $b\leq $ 4 fm, corresponding to the FOPI 
centrality cut. The $K^+$ rapidity distributions $dN/dY^{0}$ 
($Y^{0} = Y_{lab}/Y_{CM}$ -1) for this reaction is shown  
in Fig.\ref{dndy_fig}. Again the 
BRP in-medium kaon potential has been used. 
Consistent with our previous results  \cite{fuchs01} and those from 
other groups \cite{nantes02}, this figure shows that the existence of 
a repulsive kaon potential is required to match the experimental 
data \cite{menzel00,fopi97}, in particular around mid-rapidity. 
The influence of the nuclear EOS is thereby on the 20\% level. 
The best agreement with the data  is obtained with 
the soft EOS and including a kaon potential.   

The next figure gives the corresponding proton flow. 
A $p_{t}$ cut of $p_{t}/m > $ 0.5 accounts for the 
experimental acceptance \cite{fopi95,fopi99}. 
At this high energy the dependence of the 
proton flow on the nuclear EOS is weak and both options, 
i.e. a soft and a hard EOS reproduce the FOPI data \cite{fopi95} 
reasonably well. It should be noted that the analysis has been 
performed for all protons. At large rapidities the experimental 
flow is rather well reproduced whereas deviations occure between 
$ -1\leq Y^{0}\leq 0$ where the consideration of cluster effects 
might help to improve on the data fitting. It is, on the other hand, 
known from Ref. \cite{nantes97} that proton and kaon flow are 
only very loosely connected. The kaon producing sources 
carry this large in-plane flow. The kaons themselves carry a 
much smaller flow fraction since, at given 
rapidity, they are produced from two baryons originating from 
very different rapidity regions. 
Therefore baryon sources with positive and negative 
$p_x /m$ values add up to an almost vanishing net flow 
(see the results given by open down triangles 
in Fig.\ref{pflow2_fig}). $\Lambda$'s which are produced in association 
with $K^+$'s have very similar flow pattern as protons, i.e. 
they show almost the same $p_x/m$ scaling \cite{fopi99}. 
In \cite{wang99} the $\Lambda$ flow was 
investigated within the present model and the data were 
best reproduced including the  hyperon mean field
according to  SU(3) scaling 
$U_{\rm opt}^Y = \frac{2}{3} U_{\rm opt}^B$. As 
also observed in \cite{li96} the primordial $\Lambda$ flow is 
moderate but strongly enhanced by the $\Lambda N$ final state 
interactions, i.e. rescattering and the mean field. 

Figure \ref{Fig. 5} shows now the $K^+$ transverse flow 
as a function of the scaled rapidity in 1.93 AGeV 
$^{58}$Ni + $^{58}$Ni reactions. In this 
figure the full squares represent the '95 data set from 
FOPI (old data) \cite{fopi95}. The full circles show the 
'99 data from FOPI \cite{fopi99} with improved statistics, 
open circles indicate their reflection at mid-rapidity. 
The theoretical results are given by the BRP 
with a soft EOS. Around mid-rapidity the 
two calculations with $U_{K} \& $ LF (static 
potential + Lorentz force, open down triangles) and 
without $U_{K}$ (open up triangles) almost coincide. 
The result without $U_{K}$, i.e. only 
including the kaon rescattering effect, predicts 
a slightly positive flow. The result with $U_{K} \& $ 
LF leads to a very small anti-flow. Around mid-rapidity 
both calculations agree with the data within error bars. 
However, at spectator rapidities the two results with 
$U_{K} \& $ LF and without $U_{K}$ differ substantially 
from each other. With respect to the old data set both 
calculations, i.e. with $U_{K} \& $ LF and without $U_{K}$, 
agree with experiment within error bars since both 
reproduce the nearly vanishing side flow signal of 
$K^+$s around mid-rapidity. This means that 
the old data can be reproduced without need for 
an in-medium kaon potential \cite{nantes97}. 
However, the new data with much small error bars are only 
in agreement with those calculations which treat the 
kaon potential in the covariant kaon dynamics. 
This indicates that it is necessary to include the 
in-medium kaon potential in order to describe the new 
data for the $K^+$ transverse flow. 
 
From Fig. \ref{Fig. 5} it is seen that the strongly repulsive 
static potential tends to push the kaons 
dramatically away from the spectator matter, leading to a 
strong anti-flow around midrapidity (stars). 
The effect of the LF contribution in the covariant 
kaon dynamics pulls the kaons back to 
the spectator matter, resulting in a finally reasonable 
pattern of the $K^+$ transverse flow, which is in 
good agreement with the FOPI data. This feature 
of the LF contribution can also be seen 
in Fig. \ref{Fig. 6} where the calculations are performed 
by the BRP with a hard EOS. This illustrates that 
the LF like contribution, originating 
from spatial components of the vector field, provides 
an important contribution to the in-medium kaon 
dynamics in heavy ion collisions. Kaons are produced 
in the early phase of the reaction where the relative 
velocity of projectile and target matter is large. 
Thus the kaons feel a non-vanishing baryon current 
in the spectator region, in particular in non-central 
collisions. This contribution dramatically 
counterbalances the influence of the repulsive 
potential on the $K^+$ transverse flow. 

\begin{figure}
\begin{center}
\leavevmode
\epsfxsize = 15cm
\epsffile[0 50 600 700 ]{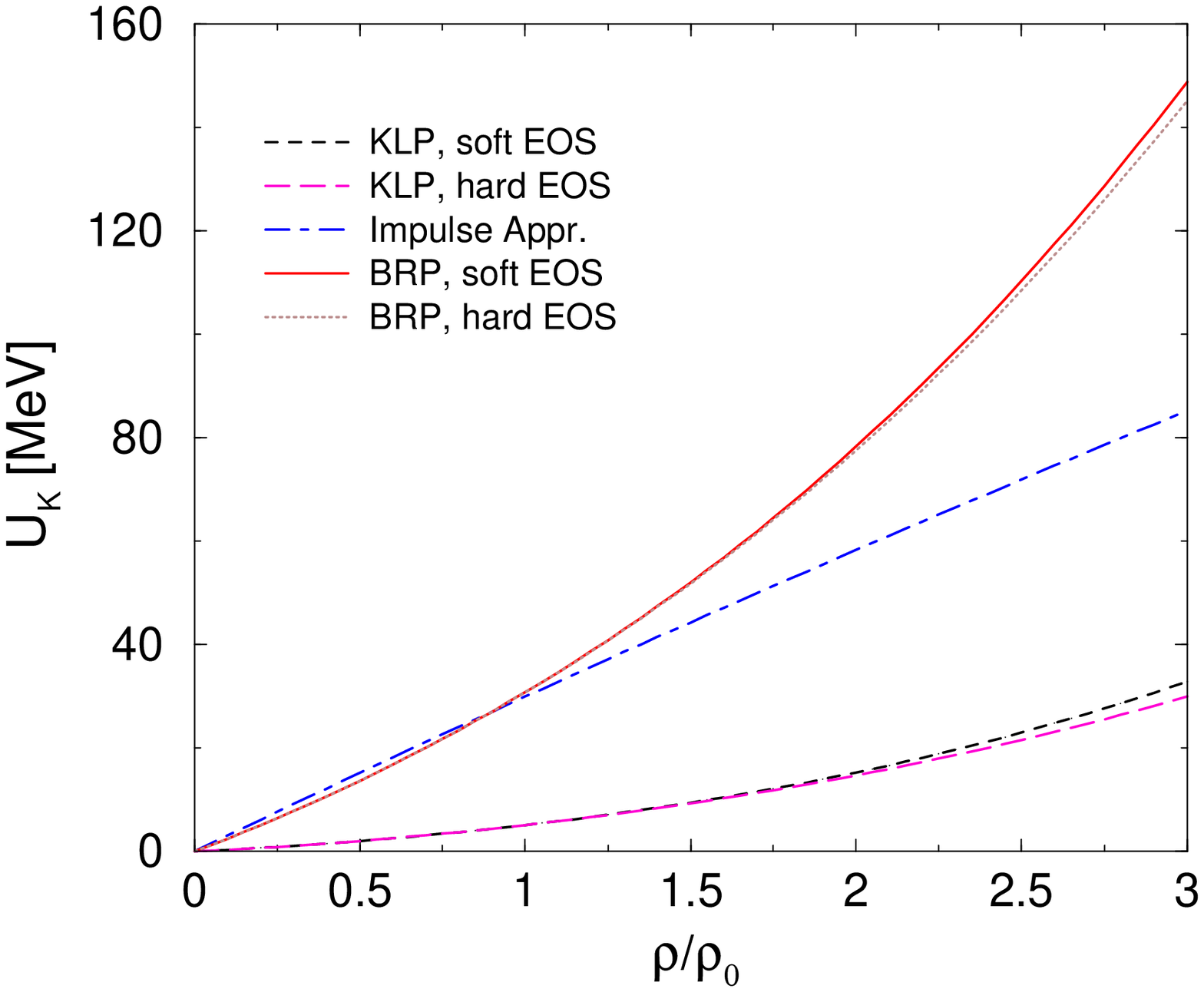}
\end{center}
\caption{Density dependence of the in-medium kaon potential 
 at zero momentum. The KLP and BRP stand for the Ko$\& $Li 
 parameterization and Brown$\& $Rho parameterization, 
respectively. }
\label{Fig. 1}
\end{figure}

\begin{figure}
\begin{center}
\leavevmode
\epsfxsize = 13cm
\epsffile[60 50 480 400 ]{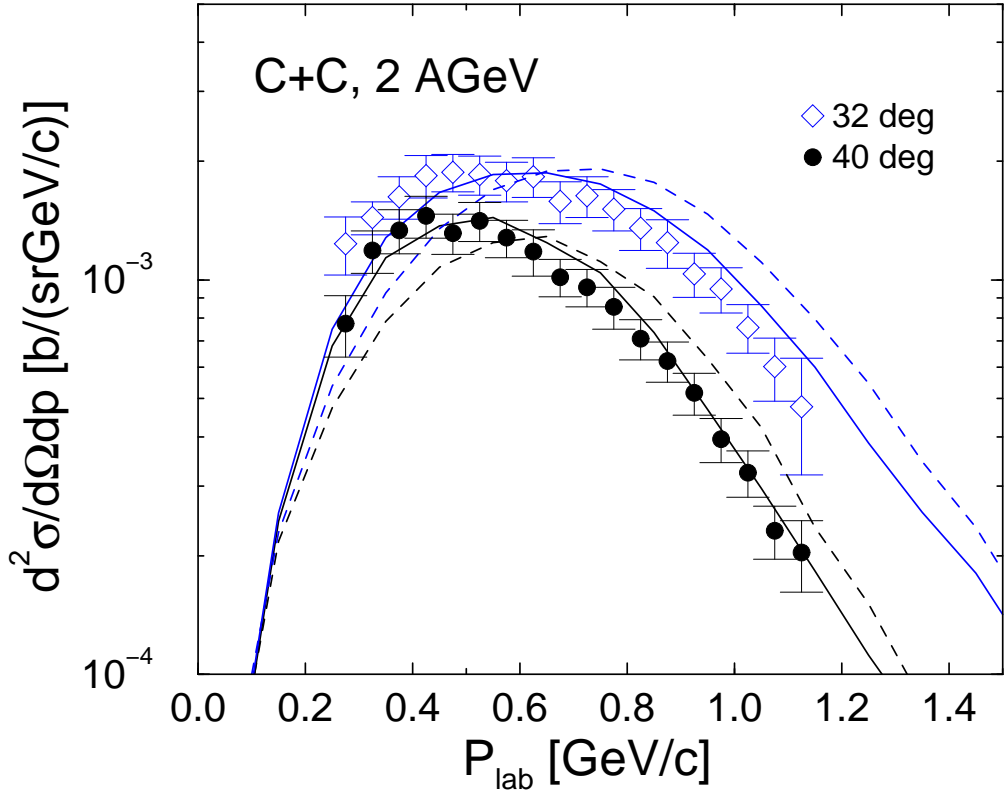}
\end{center}
\caption{Inclusive $K^+$ spectra at $\theta_{\rm lab}=32^o,~40^o$ 
in $^{12}$C +$^{12}$C reactions  
at 2.0 AGeV. The calculations are performed using different 
descriptions of the $BB\rightarrow BYK^+$ final state 3-body 
phase space. The dotted curves refer to an isotropic 3-body 
phase space while the solid curves are obtained using the 
parameterization of Eq. (19) with an additional empirical 
c.m. angular anisotropy (see text). The corresponding KaoS data are 
taken from \protect\cite{sturm02}.}  
\label{spec_fig}
\end{figure}

\begin{figure}
\begin{center}
\leavevmode
\epsfxsize = 15cm
\epsffile[0 50 500 600 ]{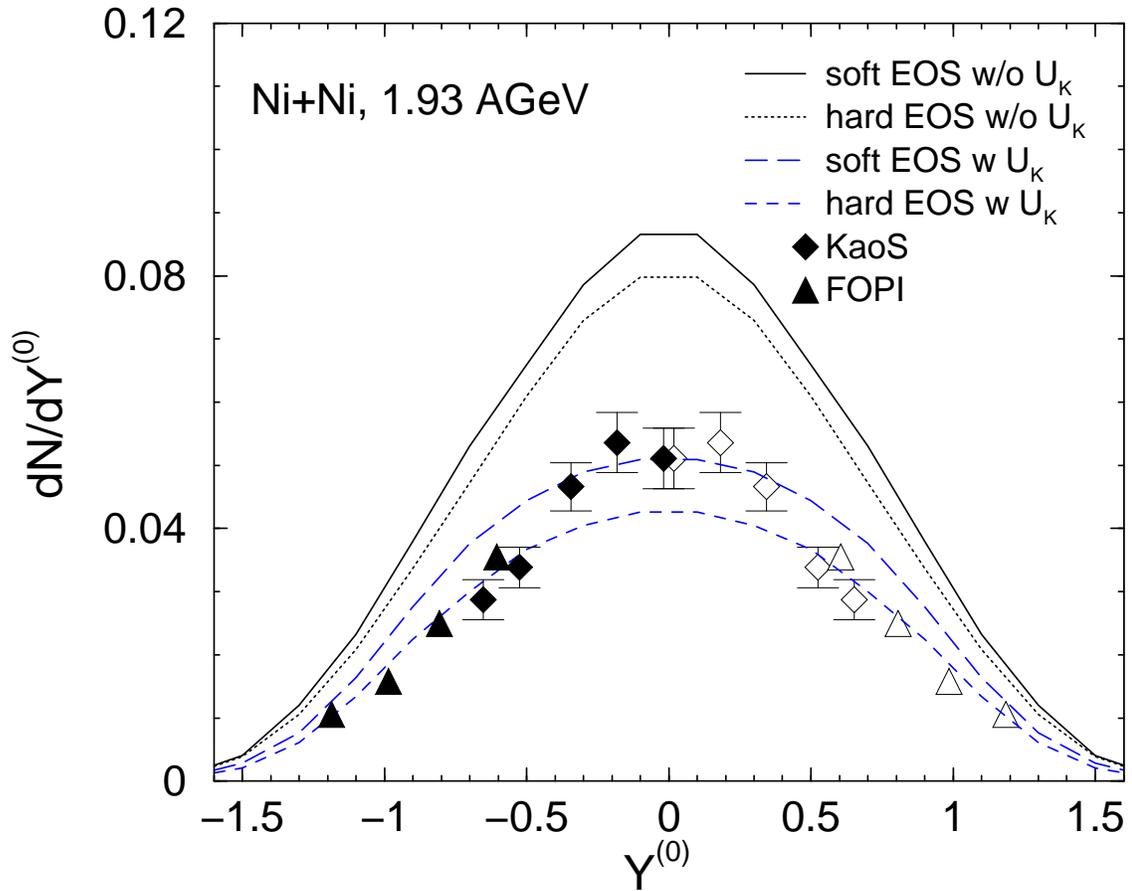}
\end{center}
\caption{ $K^+$ rapidity distributions 
in 1.93AGeV $^{58}$Ni + $^{58}$Ni reactions 
 at impact parameter b$\leq 4$ fm. 
 The calculations are performed with and without 
the BRP $K^+$ in-medium potential and using a soft/hard EOS.  
 Diamonds represent data from KaoS \protect\cite{menzel00}, 
 triangles those from FOPI \protect\cite{fopi97}. 
}
\label{dndy_fig}
\end{figure}

From the comparison between the results in Figs. \ref{Fig. 5} 
and \ref{Fig. 6}, on the other hand, it is found that the 
$K^+$ transverse flow calculated with a hard EOS gives 
rise to a slightly positive flow around 
mid-rapidity and a worse fit to the new data in 
the target region. This is due to the fact that the kaon 
in-medium potential calculated with a hard EOS is weaker 
than that one given with a soft EOS (see Fig.\ref{Fig. 1}). 
However, the $K^+$ transverse flow given with a hard EOS 
can also roughly describe the new data set within 
error bars. Thus it appears to be impossible to extract 
information on the nuclear EOS 
from the analysis of the $K^+$ side flow.

The $K^+$ transverse flow given by the KLP with a soft 
EOS is shown in Fig. \ref{Fig. 7} where the results by 
the BRP with $U_K \&$ LF for the soft EOS (open down 
triangles) are shown as well. In this figure the 
experimental data and their reflections with respect 
to mid-rapidity are represented by the same symbols 
as in Fig. \ref{Fig. 5}. The results (stars) are calculated 
by the KLP with $U_K$ but without (w/o) LF. 
The standard treatment with a static potential, 
Eq. (\ref{pot}), shows an anti-flow around 
mid-rapidity which is larger than that one given by 
the BRP with $U_K \&$ LF and is in better agreement 
with the old data. The situation changes, however, 
dramatically when the full Lorentz structure of 
the mean field is taken into account (diamonds). 
The influence of the repulsive potential of $K^+$s 
on the in-plane flow is almost completely 
counterbalanced by the velocity dependent part of 
the interaction, i.e. the LF contribution. Hence, 
no net effect of the potential is any more 
visuable. This is clearly seen from the comparison 
between this result (diamonds) and the ones 
(open up triangles) obtained w/o $U_K$. 
This seems to indicate that the static potential 
predicted by the KLP is too weak to overcome the 
cancellation effects of the LF on the flow. 
Using the BRP, the kaon in-medium potential has 
a stronger repulsive character which gives 
rise to a reasonably anti-correlated flow signal  
and leads to a  good agreement with both, the 
old and new data set from FOPI. 

In this context one has to keep in mind that 
the net source flow for the kaon production is experimentally not 
accessible. Here one has to rely on the predictions from transport
models. The analysis of the small kaon flow signal which arises to 
large amount from the cancellation of large source flow components 
carries inherent uncertainties due to experimental and 
theoretical error bars of the source flow. Investigating  
the influence of the in-medium kaon potential 
we were confronted with two effects: 
If the kaon dynamics 
is described by the standard treatment with a static 
potential, i. e. neglecting 
the effect of a Lorentz force from spatial component of 
the vector field, the calculated $K^{+}$ in-plane flow is 
not able to reproduce the new FOPI data with reasonable 
precision. The Lorentz force effect is large and it changes the $K^{+}$ 
flow patter qualitatively. We consider this qualitative result therefore as 
stable. Quantitative statement on the strength of the kaon 
potential have to be made more carefully. Since differences are 
most pronounced at large spectator rapidities where the experimental 
proton flow is  within error bars well reproduced by the present 
transport calculations, the $K^{+}$ flow 
can help to distinguish between different models. 
However, due to the smallness of the signal and present theoretical and
experimental error bars, definite quantitative statements require 
further investigations. 

\section{Conclusions}
In summary, the influence of the chiral mean field 
on the transverse flow of kaons in heavy ion 
collisions at SIS energy has been investigated 
within covariant dynamics. The kaons 
inside the nuclear medium are described as 
quasi-particles carrying effective 
masses and momenta. This accounts for the correct 
mass-shell properties of the particles inside the 
nuclear medium. 
A consequence of the covariant kaon 
dynamics is the appearance of a momentum 
dependent force proportional to the spatial components 
of the vector field which resembles the Lorentz force 
in electrodynamics. Although such Lorentz forces vanish 
in equilibrated nuclear matter they provide an essential 
contribution to the dynamics in the case of energetic 
heavy ion collisions. The influence of in-medium effects 
on the in-plane flow is counterbalanced to large extent 
by this additional contribution. 

In the present studies 
two types of in-medium potentials with different 
parameterizations have been applied. Their influence 
on the $K^+$ in-plane flow has been discussed. Our 
theoretical results show that in the covariant 
dynamics the new FOPI data can be reasonably described 
using a parameterization proposed by Brown $\&$ Rho which 
partially accounts for higher order corrections 
in the chiral expansion. The reproduction of the 
more recent FOPI data, in particular at spectator rapidities, 
requires a relatively strong repulsive $K^+$ potential, which 
is in good agreement with the ones determined from the 
kaon-nucleon scattering length using the impulse 
approximation\cite{brown94}  and also with recent information 
obtained from $p+A$ reactions \cite{anke}. This 
finding is consistent with the description of the  $K^+$ 
multiplicities in $A+A$ reactions. Most 
transport simulations reproduce corresponding data \cite{kaos} 
only when in-medium potentials are included 
\cite{ko95,giessen,nantes02,fuchs01}. In summary, information extracted 
from the analysis of the $K^+$ transverse flow is consistent 
with the knowledge from other sources.  \\

\noindent{\bf Acknowledgments}

One of the authors (Y. M. Z.) is grateful to C. M. Ko for 
useful discussions. This work is supported in part by the 
National Natural Science Foundation of China (NSFC) under 
Grant No. 19975074 and 10275096, by the Deutsche Forschungsgemeinschaft 
(DFG) under Grant No. 446CHV-113/91/1, and by the National 
Research Concil of Thailand (NRCT) under Grant No. 1.CH7/2545. 


\begin{figure}
\begin{center}
\leavevmode
\epsfxsize = 15cm
\epsffile[0 50 500 600 ]{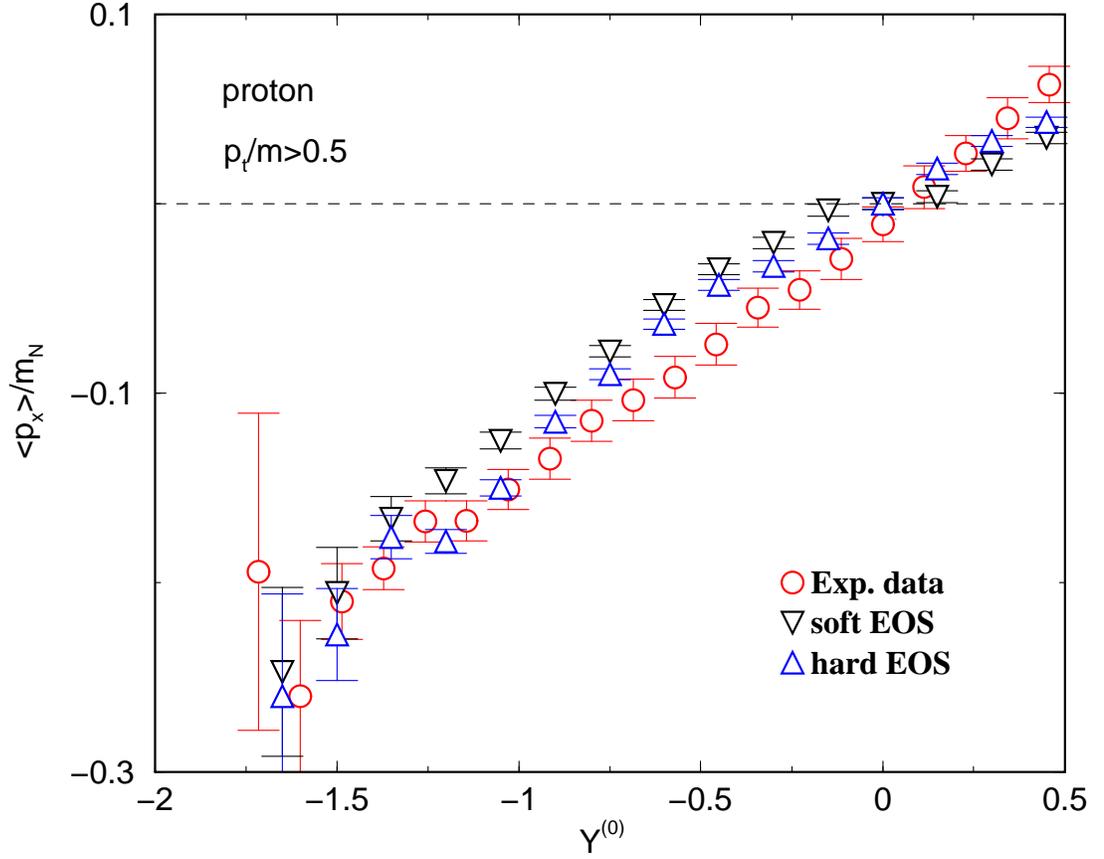}
\end{center}
\caption{Transverse proton flow as a function 
 of rapidity $Y^{0}$ 
 in 1.93AGeV $^{58}$Ni + $^{58}$Ni reactions 
 at impact parameter b$\leq 4$ fm. 
 The calculations with a soft ($\bigtriangledown $) 
 and hard ($\bigtriangleup $) nuclear EOS 
 are compared to the FOPI data \protect\cite{fopi95}.}  
\label{pflow_fig}
\end{figure}
\begin{figure}
\begin{center}
\leavevmode
\epsfxsize = 15cm
\epsffile[0 50 500 600 ]{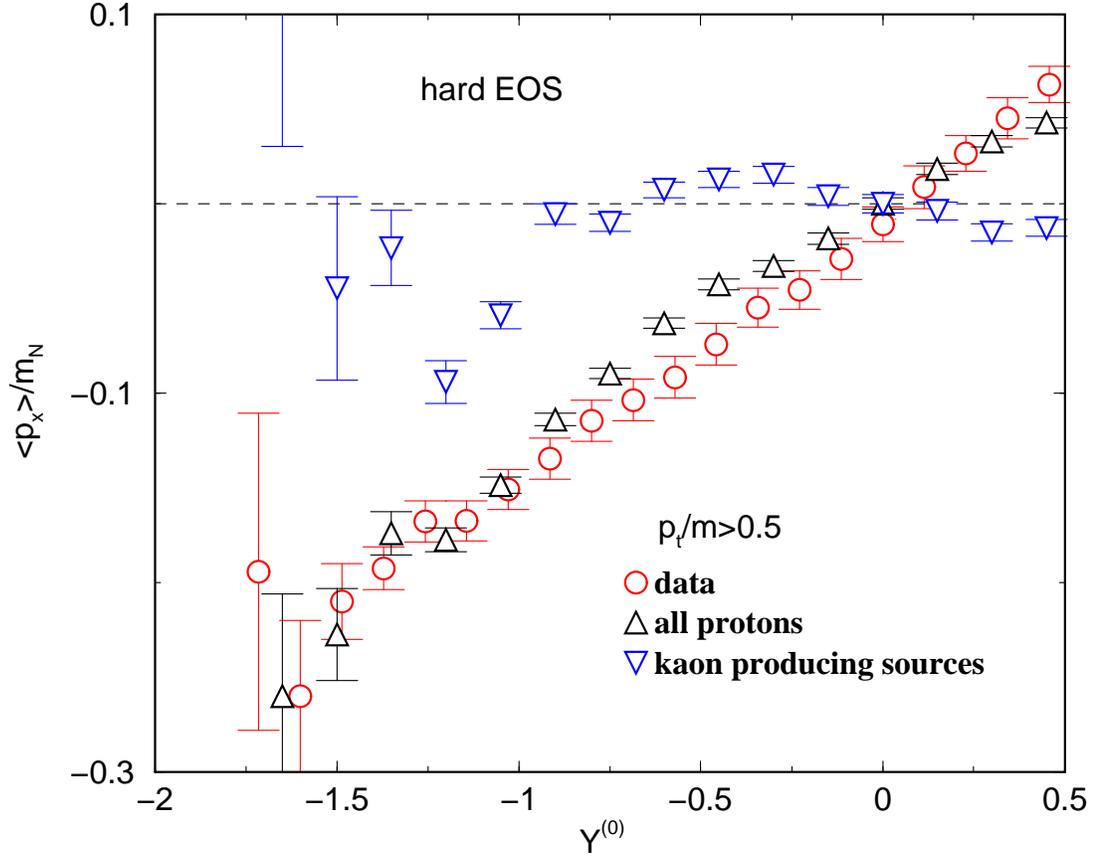}
\end{center}
\caption{Transverse flow of the kaon producing sources as a function 
 of rapidity $Y^{0}$ 
 in 1.93AGeV $^{58}$Ni + $^{58}$Ni reactions 
 at impact parameter b$\leq 4$ fm. 
 The calculations are performed with the BRP $K^{+}$ in-medium 
 potential and using a hard nuclear EOS. The FOPI data  are 
 from  \protect\cite{fopi95}.}  
\label{pflow2_fig}
\end{figure}
\begin{figure}
\begin{center}
\leavevmode
\epsfxsize = 15cm
\epsffile[0 50 500 600 ]{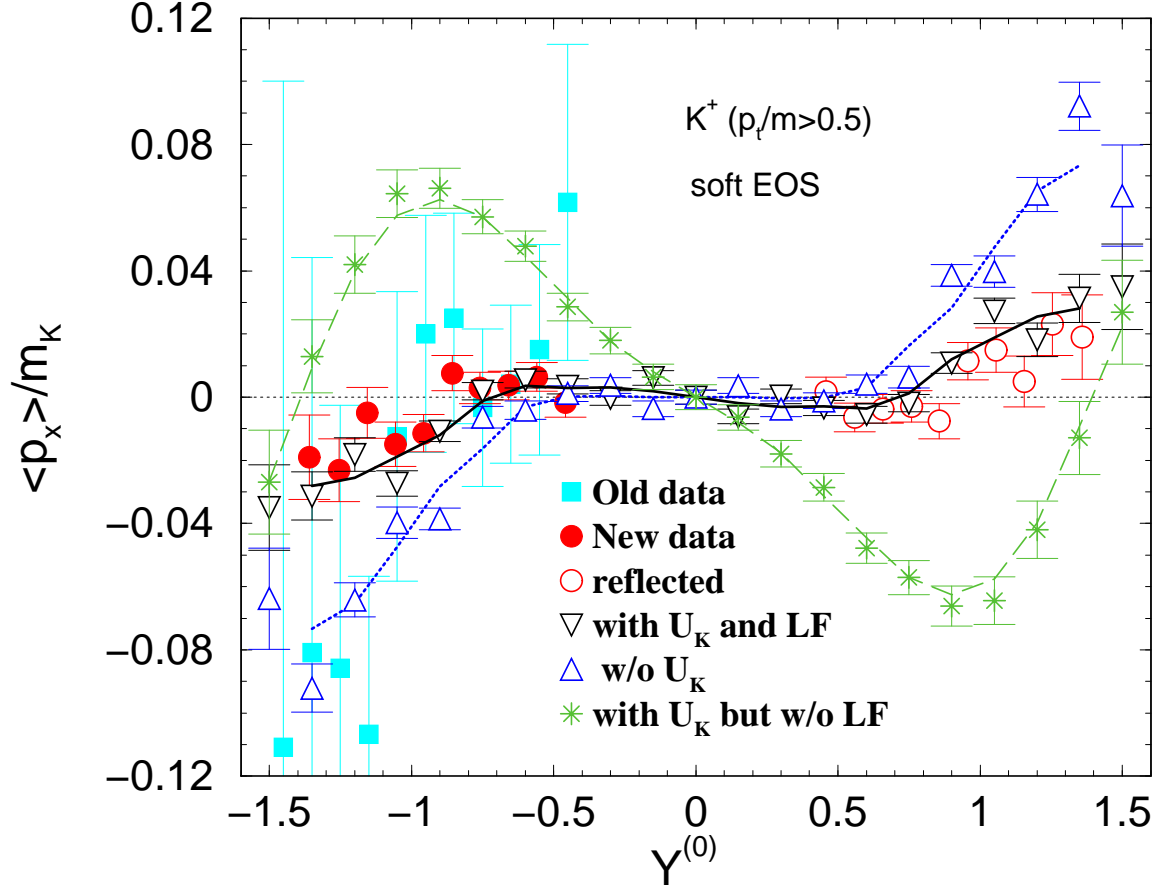}
\end{center}
\caption{ The $K^+$ transverse flow as a function 
 of rapidity $Y^{0}$ in 1.93AGeV $^{58}$Ni + $^{58}$Ni reactions 
 at impact parameter b$\leq 4$ fm. 
 The calculations are performed using the BRP $K^+$ 
in-medium potential with a soft EOS.  
 The full squares represent the 
 FOPI data from \protect\cite{fopi95}, full circles are 
 more recent  FOPI data \protect\cite{fopi99}, 
 their reflections with respect to mid-rapidity 
 are shown by the open circles. The open down 
 triangles denote the calculated results with 
 $U_K$ and the Lorentz-force (LF) contribution. 
 The open up triangles indicate the results without 
 $U_K$. The stars stand for the results with $U_K$ 
 but without LF. }
\label{Fig. 5}
\end{figure}
\begin{figure}
\begin{center}
\leavevmode
\epsfxsize = 15cm
\epsffile[0 50 500 600 ]{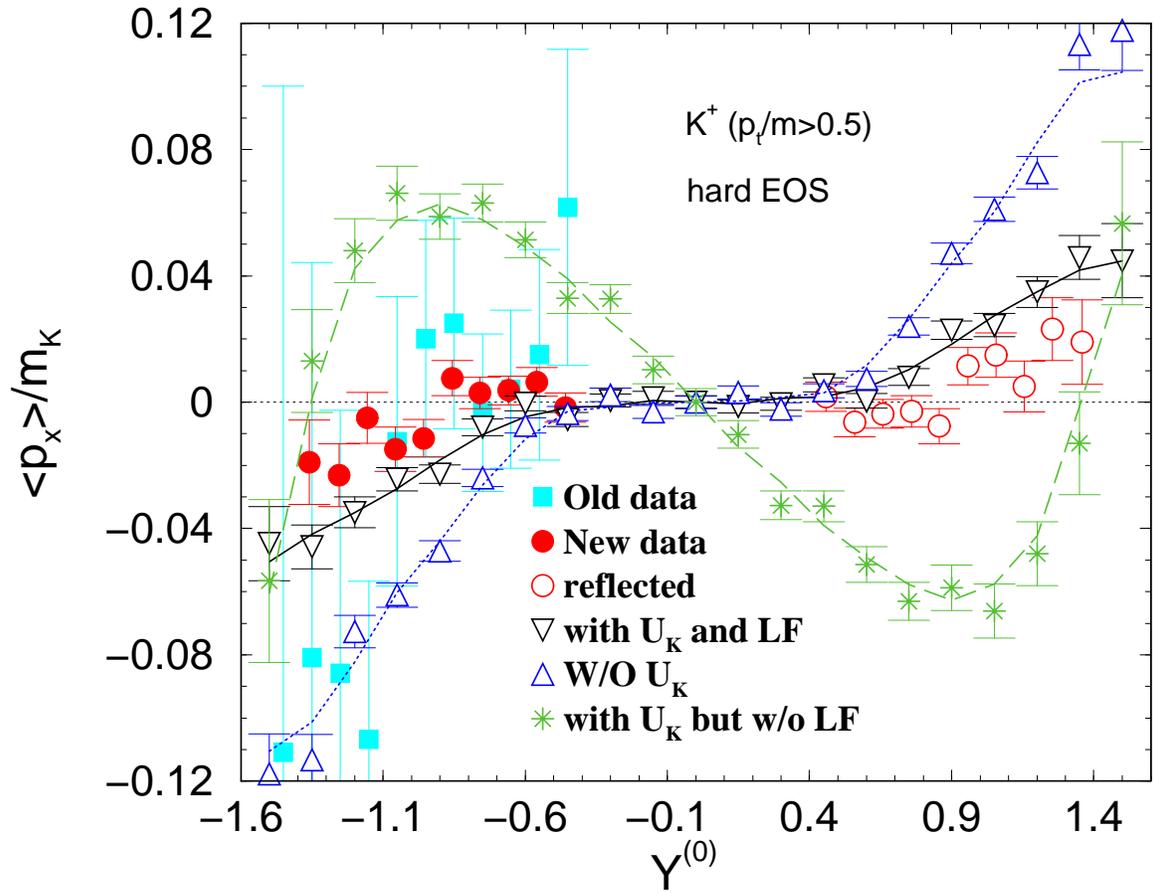}
\end{center}
\caption{ Same as Fig. \ref{Fig. 5}, but with a hard EOS. }
\label{Fig. 6}
\end{figure}
\begin{figure}
\begin{center}
\leavevmode
\epsfxsize = 15cm
\epsffile[0 50 500 600 ]{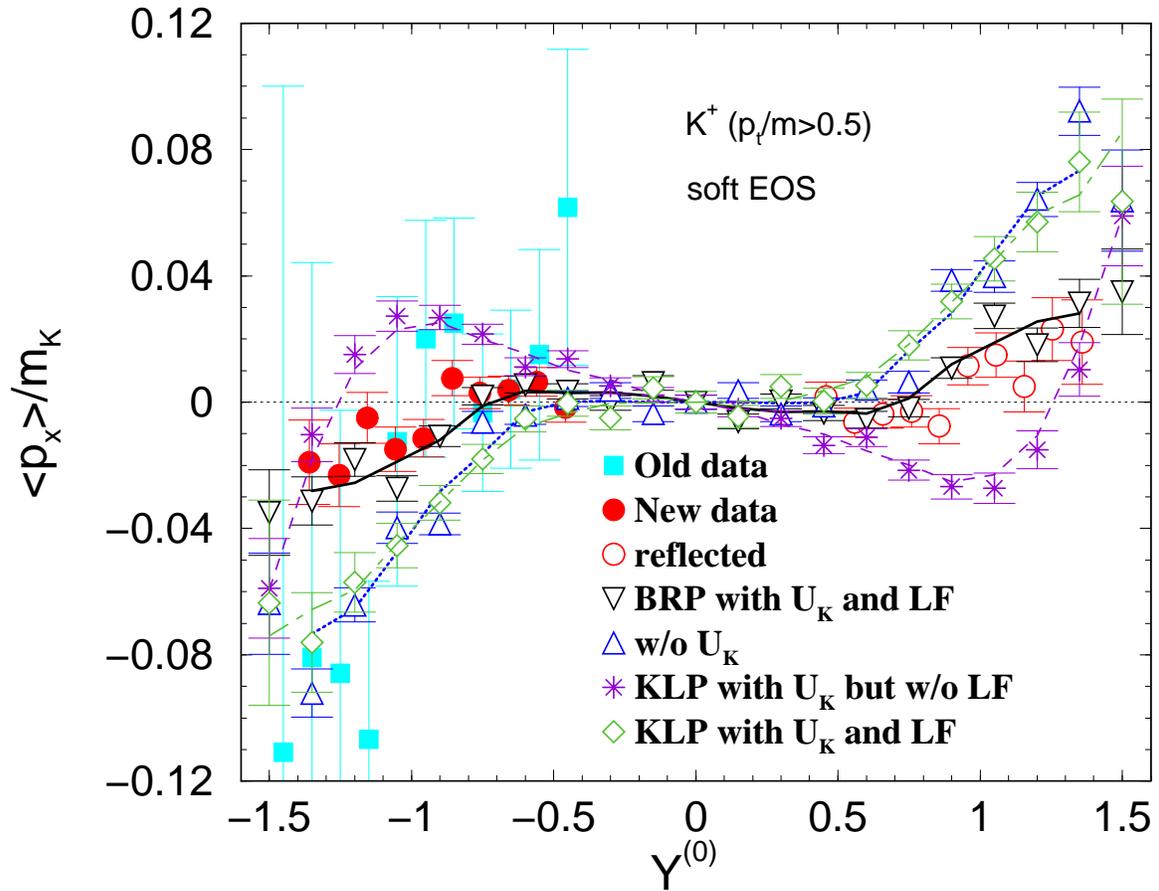}
\end{center}
\caption{Same as Fig. \ref{Fig. 5}, but by the KLP with a soft EOS.  
 Here the results by the BRP with $U_K \&$ LF are 
 also shown. }  
\label{Fig. 7}
\end{figure}
\end{document}